\documentclass[aps,pra,twocolumn,amsmath,amssymb,superscriptaddress,showpacs,longbibliography]{revtex4-1}

\usepackage{amsfonts}
\usepackage{amsmath}
\usepackage{amssymb}
\usepackage{amsthm}
\usepackage{fontenc}
\usepackage{graphicx}
\usepackage{xcolor}
\usepackage{textcomp}
\usepackage{epstopdf}
\usepackage{braket}
\usepackage{mathtools}
\usepackage{amsmath}
\usepackage{dcolumn}
\usepackage{multirow}
\usepackage{units}
\usepackage{bbm}
\usepackage{appendix}
\usepackage[normalem]{ulem}

\begin{document}

\title{One-dimensional moir\'e superlattices and flat bands in collapsed chiral carbon nanotubes}

\author{Olga Arroyo-Gasc\'on}
\affiliation{Instituto de Ciencia de Materiales de Madrid, Consejo Superior de Investigaciones Cient\'{\i}ficas, C/ Sor Juana In\'es de la Cruz 3, 28049 Madrid, Spain}
\author{Ricardo Fern\'andez-Perea}
\affiliation{Instituto de Estructura de la Materia, Consejo Superior de Investigaciones Cient{\'\i}ficas,
Serrano 123, E-28006 Madrid, Spain}
\author{Eric Su\'arez Morell}
\affiliation{Departamento de F\'isica, Universidad T\'ecnica Federico Santa Mar\'ia, Casilla 110-V, Valpara\'iso, Chile}
\author{Carlos Cabrillo}
\affiliation{Instituto de Estructura de la Materia, Consejo Superior de Investigaciones Cient{\'\i}ficas,
Serrano 123, E-28006 Madrid, Spain}
\author{Leonor Chico}
\email{leonor.chico@icmm.csic.es}
\affiliation{Instituto de Ciencia de Materiales de Madrid, Consejo Superior de Investigaciones Cient\'{\i}ficas, C/ Sor Juana In\'es de la Cruz 3, 28049 Madrid, Spain}

\begin{abstract}

We demonstrate that one-dimensional moir\'e patterns, 
analogous to those found in twisted bilayer graphene, can arise in collapsed chiral carbon nanotubes.
Resorting to a combination of approaches, namely, molecular dynamics to obtain the relaxed geometries and tight-binding calculations validated against ab initio modeling, we find that magic angle physics occur in collapsed carbon nanotubes. Velocity reduction, flat bands and localization in AA regions with diminishing moir\'e angle are revealed, showing a magic angle close to 1$^{\rm o}$.
From the spatial extension of the AA regions and the width of the flat bands, we estimate that many-body interactions in these systems are stronger than in twisted bilayer graphene. 
Chiral collapsed carbon nanotubes stand out as promising candidates to explore 
many-body 
effects and superconductivity in low dimensions, emerging as the one-dimensional analogues of twisted bilayer graphene. 

\end{abstract}

\maketitle


The discovery of superconducting \cite{Cao2018} and correlated insulating behavior \cite{Cao2018a} in twisted bilayer graphene (TBG) has shaken up the field of two-dimensional (2D) materials, reinvigorating the study of graphene-based systems. 
If two coupled graphene sheets are rotated by an angle of 1.1$^{\rm o}$ (dubbed magic angle), starting from a symmetric stacking,
the Fermi velocity drops to zero and flat bands appear at the neutrality point, 
giving rise to new physics \cite{LopesDosSantos2007,SuarezMorell2010,TramblyDeLaissardiere2010,Bistritzer2011}. 
Strong correlations near the magic angle could be foreseen due to the existence of these flat bands, producing a many-particle gap, but superconductivity in TBG was an unexpected phenomenon that awaits a theoretical explanation and promises to be a fundamental piece in the understanding of unconventional superconductors.

If the rotation angle is roughly below 10$^{\rm o}$, moir\'e patterns start to be detectable by visual inspection. Regions with AB, BA and AA stacking can be identified. 
For rotation angles under 5$^{\rm o}$ band velocity reduction is noticeable, being greatly diminished around 2$^{\rm o}$ \cite{TramblyDeLaissardiere2010,SuarezMorell2010}.     
Dispersionless bands, with zero velocity, correspond to spatially localized states. In the case of the moir\'e potential, localization in TBG 
takes place in regions with direct or AA stacking, which present a great increase of the local density of states \cite{TramblyDeLaissardiere2010}. 
These high-density regions are the origin of the exotic behavior observed near the magic-angle.  

Correlations are known to be enhanced in low dimensions; they occur in 2D systems such as bilayer graphene and even more strongly in one-dimensional (1D) systems, in which exotic many-body physics has been studied \cite{MattisBook}. For instance, in carbon nanotubes (CNT), Luttinger liquid behavior was predicted \cite{Egger1997,Kane1997} and experimentally observed \cite{Bockrath1999,Ishii2003}. If electron-electron interactions are strong, a Mott insulator behavior is expected \cite{Hubbard1963,Lieb1968,MattisBook, EsslerBook,LopezSancho2001}. It is therefore natural to look for a one-dimensional version of twisted bilayer graphene. The most immediate instance would be twisted bilayer nanoribbons. However, edge states \cite{Akhmerov2008,Jaskolski2011} are unavoidable at the terminations of TBG \cite{SuarezMorell2015} or one-dimensional moir\'e ribbons \cite{SuarezMorell2014,Pelc2015}, obscuring the role of the AA-stacked regions.

In chiral double-walled nanotubes, we can also find one-dimensional moir\'e patterns, albeit most of them are incommensurable and consequently difficult to tackle from the computational viewpoint.  Koshino {\it et al.}  theoretically found that these structures have drastic changes in their electronic properties from metallic to semiconductor \cite{Koshino2015DW}. They have been also experimentally studied but without evidence of flat bands \cite{Bonnet2016ER,Zhao2020}.
Therefore, the quest for other systems which could be the 1D analogues of TBG is of great importance to elucidate the nature of superconductivity and strong correlations recently found therein. 

Here, we propose collapsed chiral carbon nanotubes as ideal systems to explore one-dimensional moir\'e physics.  
Collapsed nanotubes were experimentally found long ago, in the early days of nanotube research \cite{Chopra1995}. 
In a narrow nanotube, with diameter $d$ smaller than 20 {\AA}, the elastic energy of the nanotube wall impedes its flattening into a ribbon shape \cite{Chopra1995,Benedict1998}. Recent experiments and simulations give diameters from 26 to  51 \AA\ as a threshold for stability \cite{Zhang2012,He2014}.  In any case, for $d> 40$ \AA\ the tubes are at least metastable in the flat form according to our calculations and the different criteria discussed by other authors \cite{Chopra1995,Benedict1998,Zhang2012,He2014}.
Thus, one-dimensional moir\'e graphene systems could be produced by squeezing wide carbon nanotubes flat. 


In order to obtain the geometries of the flattened nanotubes and their corresponding band structures and densities of states, we have devised the following strategy: Firstly, using molecular dynamics, we collapse a chiral cylindrical nanotube by applying a force between opposite sides of the tube. This force is later removed, and then the system is thermalized. If the CNT has a large enough diameter, 
it stays in the flat shape by dispersive interactions. Secondly, the tight-binding model is validated by comparison with Density Functional Theory (DFT) results for one of the smaller collapsed nanotubes, verifying that the bands near the neutrality point are well described with the semiempirical model. 
We then proceed to compute the dispersion relations and densities of states of larger collapsed nanotubes with the tight-binding approach. Details on the methods and a video of the collapse process are given in the Supporting Information.

\begin{center}
	\begin{figure*}[t]
	\includegraphics[width=1.5\columnwidth]{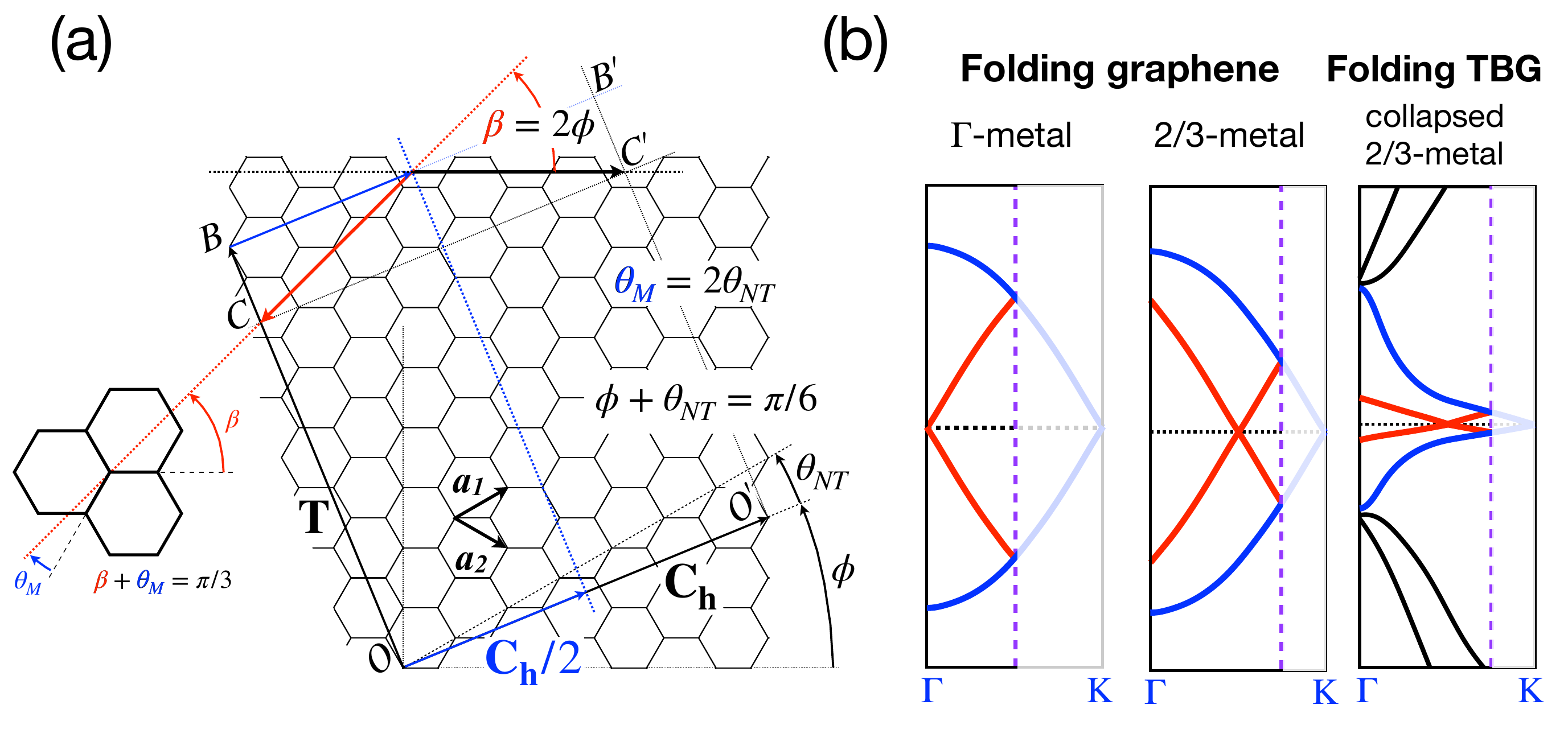}
	\caption{\small (Color online) (a) Relation between the moir\'e and the chiral angle of a collapsed carbon nanotube. 
 We take the moir\'e angle $\theta_{M}$ as the smallest angle formed between the lines defined by the sides of the hexagons.  
	The relations between the various angles indicated in the plot show that $\theta_{M}=2 \theta_{NT}$. 
	 (b) Sketch of band folding in graphene.
	 Folded bands are shown in red; the bands touching the K-point are shown in blue. The new BZ boundary is marked with a purple dashed line.  In metallic CNTs, the K-point of graphene folds either at $\Gamma$ ($\Gamma$-metals, left panel) or at 2/3 of the positive part of the Brillouin zone (2/3-metals, central panel).
	  Likewise, the right panel shows how in TBG the central pair of flattened bands fold into four central bands for a chiral, 2/3-metal, collapsed tube.}
	\label{fig:thetas}
	\end{figure*}
\end{center}

  
   Recall that CNTs are labeled with the indices $(n,m)$, which are the coordinates of the circumference vector ${\mathbf C}_h$ in a flat graphene sheet 
   (see Fig. \ref{fig:thetas}(a)) \cite{SDD1998}. 
   Chiral CNTs have $m$, $n$ nonzero and $m\neq n$; they do not possess inversion symmetry and exist in two enantiomeric forms. The chiral angle of a nanotube is that spanned between the circumference vector and the $(n,0)$ direction, as shown in  Fig. \ref{fig:thetas}(a). This angle is denoted by $\theta_{NT}$. ${\mathbf C}_h$ forms an angle $\phi$ with the horizontal direction, so $\theta_{NT} + \phi = 30^{\rm o}$. 

The chiral angle of a nanotube can be easily related to the nanotube indices $(n,m)$ \cite{SDD1998}.
 Within a simple one-orbital tight-binding model and neglecting curvature effects, an $(n,m)$ CNT is metallic if $n-m$ is a multiple of 3 and semiconductor otherwise \cite{Hamada1992,Saito1992,Saito1993}. 
The moir\'e angle of a twisted bilayer graphene is defined as the smallest relative rotation angle between the two layers, starting from a symmetric position, that we take here to be an AA stacking. We label this angle as $\theta_{M}$. Figure \ref{fig:thetas}(a) depicts that the moir\'e pattern of a collapsed chiral tube is obtained by folding the unrolled unit cell with respect to its central axis (blue dotted line). This moir\'e pattern arises due to the misalignment of the two halves of the nanotube unit cell. Rolling up the nanotube amounts to identify the points $O$ and $O'$ as well as $B$ and $B'$, which are the corners of the flattened rectangular unit cell. We take as reference the horizontal carbon bond direction, given by the black vector ending at $C'$. Upon folding, the black vector falls onto the red vector ending at $C$. Therefore, the carbon bonds of the upper and lower graphene strips form an angle $\beta$, as 
marked in the unrolled nanotube unit cell and 
in the left part of Fig.\ \ref{fig:thetas}(a), that shows a portion of the graphene lattice demonstrating that $\beta$ and $\theta_M$ add up to 60$^{\rm o}$ and are therefore equivalent. Since $\beta = 2\phi$, 
as it can be inferred from Fig.\ \ref{fig:thetas}(a), these relations lead to $\theta_{M} =2 \theta_{NT}$.  Note that rolling the flat unit cell outwards or inwards yields the two enantiomeric forms of the nanotube ($(n,m)$ vs. $(m,n)$) and, correspondingly, moiré angles with opposite sign upon flattening. 

Metallic CNTs can be classified in two types: either they have the Dirac point at $\Gamma$ or at 2/3 of the positive part of the first Brillouin zone (1BZ); they can be dubbed 2/3-metals or $\Gamma$-metals. 
Fig.\ \ref{fig:thetas}(b) schematically shows how the Dirac point of monolayer graphene is folded at $\Gamma$ (left panel) or at 2/3 of the 1BZ (central panel) for a metallic tube. In the same line of reasoning, the bands of a collapsed metallic chiral nanotube can be envisaged by folding the bands of the corresponding TBG, as shown in the right panel of Fig.\ \ref{fig:thetas}(b). This allows us to anticipate that the four central flat bands of TBG correspond to a eight-band set in collapsed chiral CNTs.

\begin{center}
	\begin{figure*}
	\includegraphics[width=1.5\columnwidth]{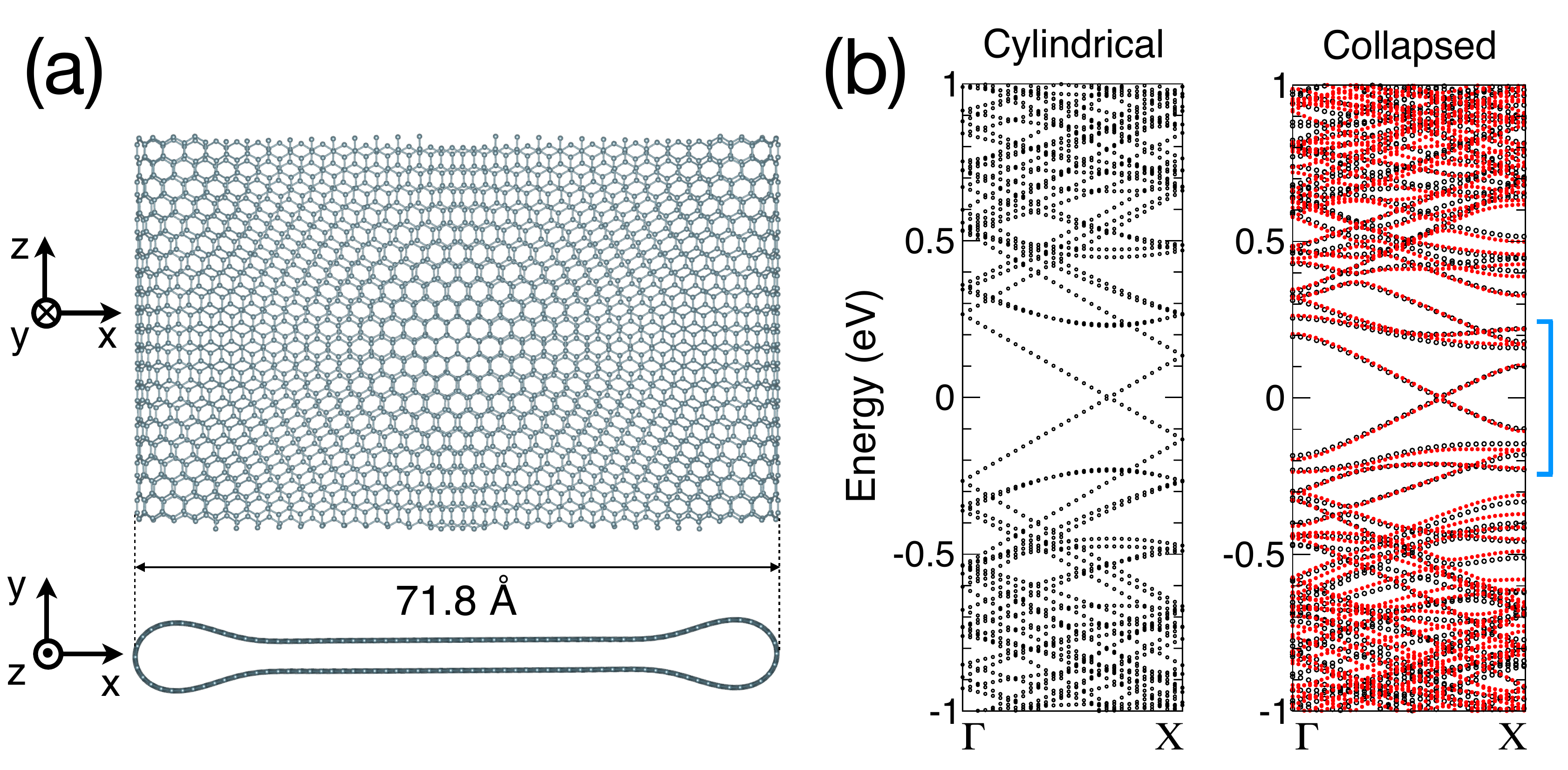}
	\caption{\small (Color online) 	(a) Unit cell of a collapsed (62,2) CNT showing our choice of axes. 
	The size of the system is indicated by an arrow.
	Upper panel: The AA stacking region is clearly seen at the center of this top view. Lower panel: Cross section of the collapsed (62,2) nanotube shows the shape of the lobes upon MD relaxation.  This CNT has 2648 atoms in the unit cell and a chiral angle $\theta_{NT}=1.58 ^{\rm o}$. In TBG, such a rotation angle yields an appreciable velocity reduction.
(b) Band structures of the (62,2) CNT 
	   ($\theta_{M}=3.16 ^{\rm o}$). Left panel: cylindrical shape with the TB approximation; right panel: collapsed geometry calculated with the TB approximation (black) and the DFT approach (red). The blue square bracket at the right marks the approximate energy span at X of the eight bands that flatten for smaller moir\'e angles. The Fermi energy is set at 0 eV. }
\label{fig:tube622}
	\end{figure*}
\end{center}

Low chiral angles can be achieved if the circumference vector is large and close to the zigzag direction ($\theta_{NT}= 0^{\rm o}$). 
 In order to obtain full moir\'e islands with AA stacking, the nanotube should have a $C_{2n}$ symmetry axis. This can be understood from Fig.\ \ref{fig:thetas}(a). The chiral vector depicted corresponds to a (6,1) CNT. As it can be seen, the two halves of the unit cell which are folded to produce the moiré pattern are not graphene supercells, so they cannot generate a TBG. To do so, the vector ${\mathbf C}_h/2$ should join two atoms of the graphene lattice. This happens if the nanotube unit cell has a $C_{2n}$ symmetry axis. 
We have chosen nanotubes with indices $(2m,2)$, i.e., 2/3-metals with $C_2$ symmetry which yields one-dimensional moiré patterns. Figure \ref{fig:tube622} (a) shows a top view of the unit cell and a transversal  cut of a (62,2) collapsed CNT. The AA island is clearly seen at the center.
Interestingly, rolling the unit cell in the transverse direction moves the AA region along the nanotube axis (see video of the rolling mode in the Supporting Information). 

%


In order to validate the tight-binding model for CNTs, we have performed {\it ab initio} calculations\cite{Soler2002} for a collapsed (62,2) CNT using the relaxed geometry obtained with MD simulations \cite{Plimpton1995,LAMMPS}.

Figure \ref{fig:tube622} (b) depicts the band structure of the cylindrical CNT calculated with the TB approximation (left panel), compared to the bands of the collapsed tube computed with both, TB (black) and DFT (red), approaches (right panel). Band velocity reduction due to flattening is evident; additionally, the almost perfect linear dispersions of the central bands in the cylindrical nanotube are softened upon collapse. Notice that the two central bands stemming from the Dirac cone of graphene are 
separated from the rest in the collapsed tube, with gaps at $\Gamma$ and X both for electrons and holes. 
DFT and TB calculations show a remarkable agreement, especially in the eight bands around the Fermi energy ($E_F$), validating the TB model chosen for the collapsed tubes. These eight bands are those that flatten at the magic angle and separated in energy from the rest, so it is important to describe them properly.


\begin{center}
	\begin{figure*}
	\includegraphics[width=1.8\columnwidth]{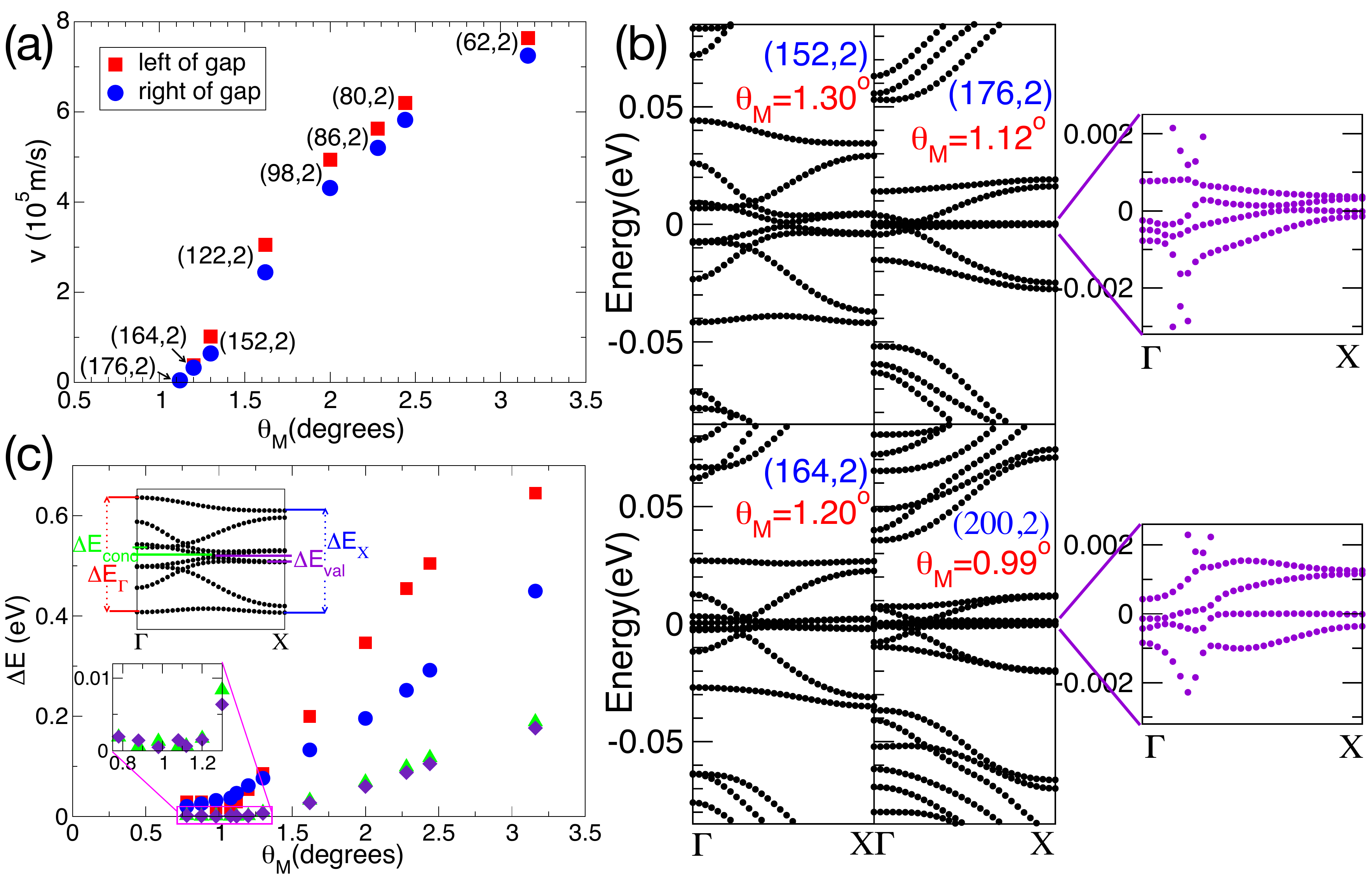}
	\caption{\small (Color online) (a) Velocities of the conduction band near the gap as functions of the moir\'e angle 
	$\theta_M$, calculated at the right (blue dots) and left (red squares) of the band minimum, as it is schematically indicated in the inset. 
   (b) Band structures of four collapsed CNTs: (152,2), (164,2), (176,2) and (200,2), calculated with the TB approximation. 
	For the (176,2) and (200,2) the four central bands are zoomed in the right panels. The Fermi level is set at zero.
	(c) Energy spans of the eight-band set at $\Gamma$ and $X$, 
		$\Delta E_{\Gamma}$ and $\Delta E_{X}$, and of the two central bands of the group, the lowest conduction and highest valence bands, 
	$\Delta E_{\rm cond}$ and $\Delta E_{\rm val}$, as functions of the chiral angle $\theta_M$. The magnitudes represented are schematically indicated in the upper inset. The lower inset zooms the $\Delta E_{\rm cond}$ and $\Delta E_{\rm val}$ data for the smaller chiral angles. }
	\label{fig:vbd}
	\end{figure*}
\end{center}
Band velocity reduction
can be quantified as a function of the chiral angle. 
As in TBG,
the velocities at the left and right of the gap opened at the Dirac cone are slightly different, as well as those corresponding to the conduction and valence bands. 
Fig.\ \ref{fig:vbd}(a) depicts the conduction band velocities of
for a series of tubes,
showing the values at the left and right of the gap as functions of the moir\'e angle.
Only values down to $\theta_{M}= 1.12^{\rm o}$, 
which corresponds to the (176,2) nanotube, are presented. For smaller angles 
the dispersions are far from linear, so near the Dirac point of folded graphene it is difficult to define a proper band velocity.

Figure \ref{fig:vbd}(b) shows how the bands close to $E_F$ evolve with decreasing chiral (and moir\'e) angle. Four cases are depicted, namely, (152,2), (164,2), (176,2) and (200,2), with $\theta_{M}$ between $1.3^{\rm o}$ and $0.98^{\rm o}$, 
close to the magic angle in TBG.
As commented before, the eight central bands are gradually isolated from the rest of the spectrum. Their energy span also narrows, with flatter dispersions in general, albeit their behavior is not so simple as in TBG.
For the two smaller chiral angles, corresponding to the (176,2) and (200,2) collapsed CNTs, we zoom the central bands, with extremely flat portions, but also revealing anticrossings and complicated dispersions. 
Such behavior is related to the folding of the central flat bands; level repulsion leads to the anticrossings. 
These two large nanotubes result to be metallic within the one-electron approach, as it is evident from the zoom. However, as we comment below, a gap should open if many-body interactions are included. The Fermi level crosses the two central bands; their energy widths are smaller than 1 meV for the two bands of the (176,2) collapsed tube and for one of the (200,2) collapsed tube. 


Consequently, for small chiral angles the flatness of the bands cannot be well described by band velocities. Therefore, we
assess band flatness 
  by the energy span, defined as the difference between the maximum and minimum energies of each band, irrespective of the wavevector.
In fact, this is the relevant magnitude to estimate the importance of electron-electron interactions. We focus on the two bands forming the gap in the smaller CNTs, which are generally the flattest.
We also report the full energy span of the central eight-band set. This takes place either at 
$\Gamma$ or X, depending on the nanotube: 
up to the (152,2), it is largest at $\Gamma$. In Fig.\ \ref{fig:vbd}(c) we show
 the energy difference between the highest and lowest bands of the set at $\Gamma$, $\Delta E_\Gamma$, and at X, $\Delta E_X$, as well as the energy spans of the two central bands. For simplicity, we label the latter conduction and valence bands, $\Delta E_{\rm cond}$ and $\Delta E_{\rm val}$.
Since not all of them reach their minimal values for the same tube, 
a complementary analysis is necessary. 
$\Delta E_\Gamma$ is minimum for the (176,2) case; however, $\Delta E_X$ continues decreasing with diminishing moir\'e angle. Overall, the two central bands are also flatter for the (176,2) tube. However, the absolute minima are the (182,2) tube with $\Delta E_{\rm cond}=0.57$ meV and the (200,2) tube with $\Delta E_{\rm val}=0.48$ meV.
These values are to be compared to those of the (176,2)
case,
namely $\Delta E_{\rm cond}=0.68$ meV and $\Delta E_{\rm val}=0.63$ meV. 
But notice 
 that the other band widths of the (182,2) and the (200,2) tube are above 1 meV. Besides, folding has an important effect on the band dispersions. The two central conduction (valence) bands stem from one band of the TBG. In order to compensate for dispersions arising from band repulsion, we can consider then the energy span of each folded pair. With this criterion, the minimum energy spread is attained for the (176,2) CNT, being 1.16 meV for the upper pair and 1.65 meV for the lower pair of bands of the central quartet.

\begin{center}
	\begin{figure*}[t!]
		\includegraphics[width=1.5\columnwidth]{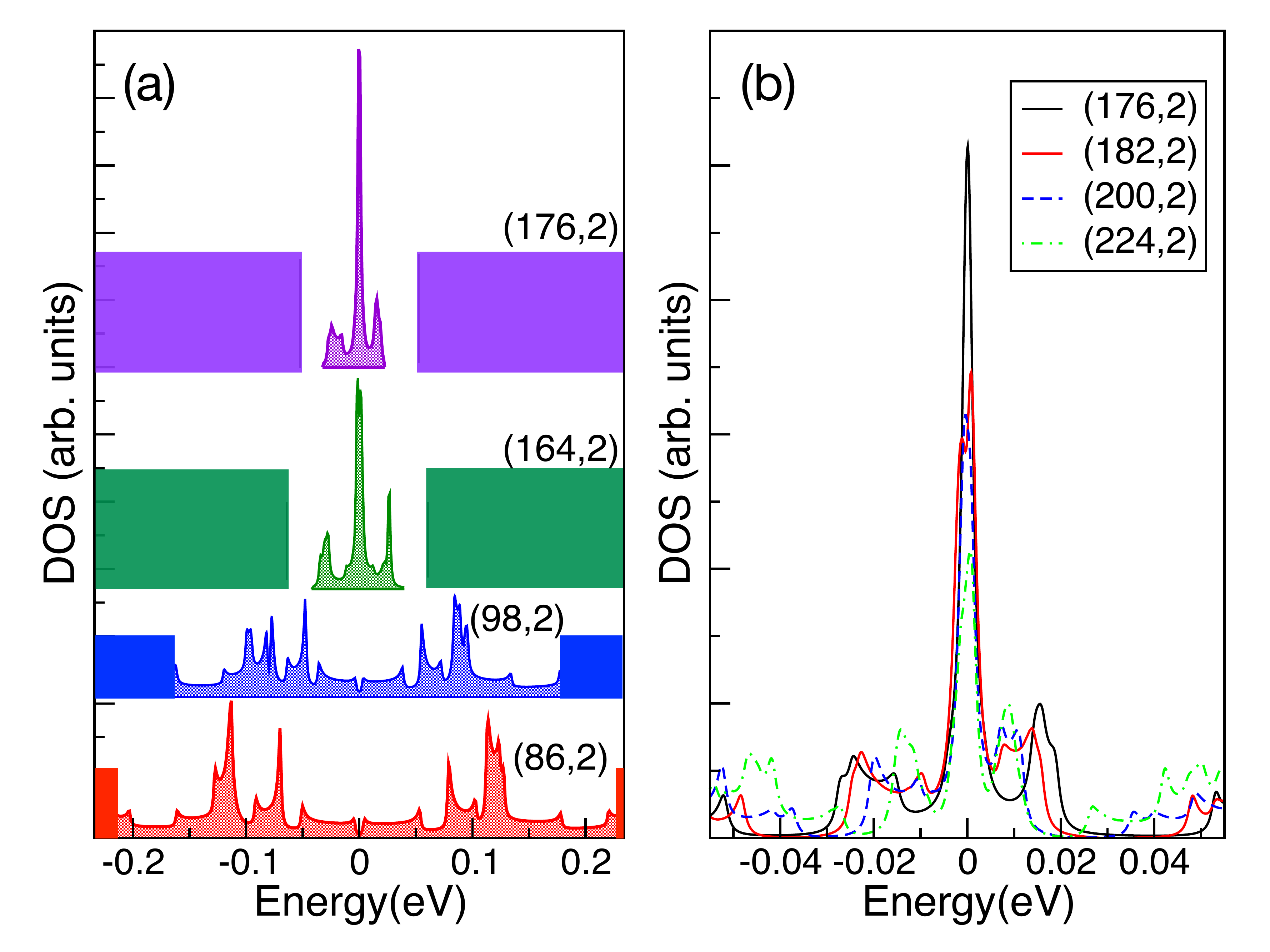}
	\caption{\small (Color online) (a) DOS for several collapsed CNTs, normalized to the number of atoms in their unit cells, calculated from the Green's function.
	The DOSs displayed correspond to the eight central bands that eventually flatten for small chiral angles. The energy edges of the upper and lower  bands surrounding the eight-band set
	are marked with colored rectangles. The DOSs are shifted maintaining the scale, so that they can be compared; the baselines correspond to their zero values. (b) DOS for the (176,2) to (224,2) tubes computed from the dispersion relations.
	$E_F$ is set at 0 eV. }
	\label{fig:severalDOS}
	\end{figure*}
\end{center}
Notwithstanding, we resort to an additional criterion to elucidate the occurrence of the magic angle: the localization of the electronic states in these bands as inferred from 
the density of states (DOS).  
For the smaller CNTs studied here, we can obtain the density of states by computing the Green's function. For larger tubes, we have to resort to the dispersion relations (see Methods in the Supporting Information). 

We present in Fig.\ \ref{fig:severalDOS}(a) 
the total DOS for several collapsed CNTs, normalized to the number of atoms in their respective unit cells, computed from the Green's function.  
Only the DOS of the eight central bands is presented; the energies of the surrounding bands are marked with colored rectangles.
The (86,2) and (98,2) collapsed tubes show similar characteristics, with a gap at $E_F$.
For the (164,2) CNT the gap disappears, and a prominent peak at $E_F$ singles out, clearly composed of several very close peaks due to the narrowness and proximity of the central bands. More conspicuously, the eight-band set is completely confined by one-particle gaps, separated from the rest of the spectrum. These features also appear in the (176,2) CNT, which shows the most prominent maximum. 
This is the largest case for which we can compute the DOS from the Green's function.
Fig.\ \ref{fig:severalDOS}(b) shows the DOS for larger collapsed CNTs computed from the dispersion relations. 
We verify that the maximum of localization occurs for the (176,2) tube, 
allowing us to identify it as the closest case to the magic angle, 
around 1.12$^{\rm o}$.

\begin{center}
	\begin{figure*}[t]
		\includegraphics[width=1.5\columnwidth]{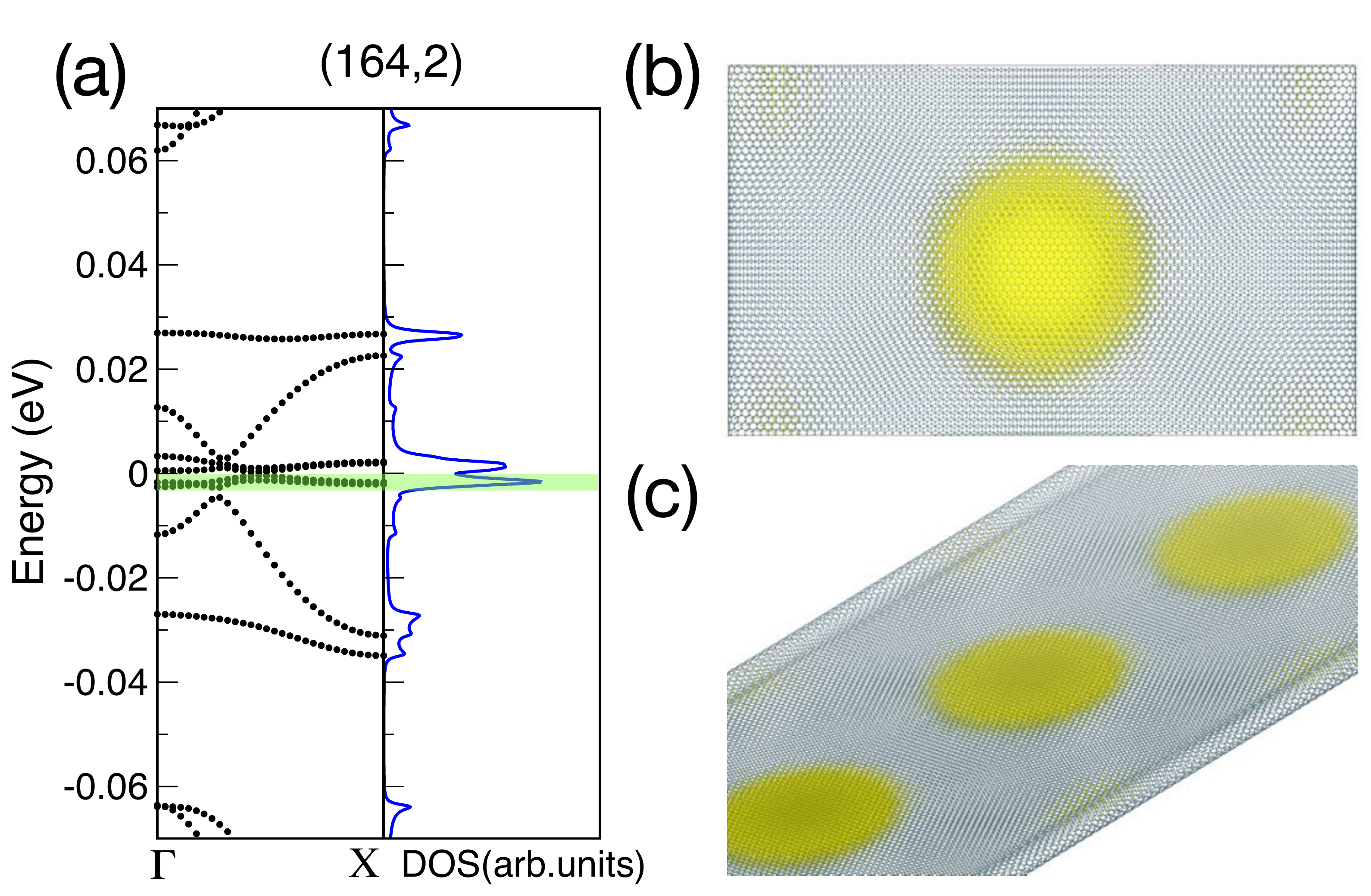}
	\caption{\small (Color online) 
	(a) Band structure and total DOS for a (164,2) collapsed nanotube. The DOS is computed from the dispersion relations with a Lorentzian broadening of 0.5 meV. The two highest valence bands are highlighted to indicate the range of energies occupied to map the spatial localization of those states. (b) Top and (c) lateral views of the structure with the LDOS  of the aforementioned bands highlighted in color. The states are localized in the AA regions of the nanotube unit cell.}
	\label{fig:ldos1642}
	\end{figure*}
\end{center}

These sharp peaks signal the spatial localization of the states. To find out where it takes place,
we map the spatial distribution of the states belonging to the two top valence bands within the unit cell. In Fig.\ \ref{fig:ldos1642}(a) the shaded area shows the range of energies used to calculate the local density of states, 
simulating a scanning tunneling microscopy image where the tip sweeps a range of energies between $E_F$ and $E_F-3$ meV. 
The states are localized in the AA region, similarly to 
 TBG.


Our numerical results permits us to quantify the strength of the Coulomb interaction in these systems, that can be mapped to a (quasi)-one-dimensional Hubbard model
in which
the effective
interacting units are the large electron concentrations in the AA regions. 
We can assess the relevance of interactions by estimating the onsite and nearest-neighbor Coulomb repulsion from the size of the AA regions of localized electrons, the CNT unit cell length and the bandwidths. 
The size of the electronic density accumulation in the AA region has a diameter of one half of the unit cell approximately (see Fig.\ \ref{fig:ldos1642}). 
Assuming one electron per band, the onsite Coulomb repulsion is $U=ke^2/R$, where $k$ is the Coulomb constant and $R$ is the radius of the AA region, so we obtain $U\simeq 500$ meV for the (176,2) CNT. If screening by a substrate is considered, this value is reduced by the corresponding dielectric constant $\epsilon$, but note that collapsed tubes can be stable in a suspended geometry, where substrate screening can be avoided. Likewise, the nearest-neighbor Coulomb repulsion is given by $V=ke^2/T$, where $T$ is the size of the unit cell and therefore the distance between charges in the 1D system.  For a (176,2) tube we have  $V \simeq 120$ meV. The energies 
$U$ and $V$ are to be compared to the bandwidth, $\Delta E \lesssim 1$ meV. The band structure allows us to extract the effective hopping parameter, $t$, given that $\Delta E = 4t$. Therefore, we  conclude that these collapsed CNTs should be strongly correlated systems, since $U/t \gg 1$ and $V/t\gg 1$, 
 near the magic angle.


In summary, we have shown that collapsed chiral carbon nanotubes behave as
quasi-one-dimensional moir\'e superlattices, being a 1D analogue of TBG. Band velocity reduction and the emergence of flat bands with localized states in AA-stacked regions also occur in collapsed CNTs. Instead of the central four-bands found in TBG, we have an eight-band set, which can be understood by folding. They are surrounded by one-particle gaps when localization emerges. We found a magic angle of 1.12$^{\rm o}$, close to the 2D value.  

The width of the central flat bands for the magic angle is significantly narrower than in TBG, being less than 1 meV. Thus, the  estimated Coulomb energies vs.\ the widths of the flat bands in collapsed tubes imply that interaction effects are much stronger in these quasi-1D systems than in TBG.

We expect our results to spur experimental research on chiral collapsed carbon nanotubes for the exploration of novel strongly correlated physics and superconductivity in low dimensions. 
Recent experiments show an impressive control on the diameters and collapse mechanisms \cite{Zhang2012,He2014}. 
It is therefore plausible to consider these systems as potential platforms for further developments in magic angle physics. 

From the theoretical viewpoint, the possibility of testing strongly correlated models and theories in a quasi-one-dimensional system is tremendously attractive. Most exactly solvable models in many-body physics are one-dimensional; providing a real material in which theoretical models can be confronted is undoubtedly appealing.

\acknowledgments

 We are very grateful to Gloria Platero for generously sharing her computational resources.
 We thank the Centro de Supercomputación de Galicia, CESGA, (www.cesga.es, Santiago de Compostela, Spain) for providing access to their supercomputing facilities.
 LC, RFP and CC acknowledge the financial support of the Spanish MCIU and AEI and the European Union under Grants No. PGC2018-097018-B-I00 (MCIU/AEI/FEDER, UE) (LC) and grant No. MAT2016-75354-P (AEI/FEDER, UE) (CC, RFP).  ESM acknowledges financial support from FONDECYT 1170921.

\section*{Supporting Information}

The Supporting Information is available free of charge at  
\url{https://pubs.acs.org/doi/10.1021/acs.nanolett.0c03091}

It includes the following files: 
\begin{itemize}
  \item ArroyoGasconACSv3-SI.pdf: PDF file describing the Models and Methods in detail. 
  \item SupportingInfo164-2.mp4: MP4 video. Lateral visualization of the (164,2) carbon nanotube collapse 
  from its cylindrical shape until the initial step of deactivation of the aligning Lennard-Jones potential. 
  \item SupportingInfo62-2rolling.mp4: MP4 video showing the rolling mode of a (62,2) nanotube. 
  The two cyan highlighted atoms serve as a guide to the eye to follow the actual displacement. 
  Note that the AA pattern moves in the axis direction. 

\end{itemize}

%

\end{document}